\documentclass{article}
\usepackage{graphicx}
\newtheorem{lemma}{Lemma}
\newtheorem{theorem}{Theorem}
\newtheorem{definition}{Definition}
\newtheorem{notation}{Notation}
 
\title{Transformation of auto-B\"{a}cklund  type for hyperbolic generalization of Burgers equation}
\author{E.V. Kutafina\\   Faculty of Applied Mathematics\\
 AGH University of Science and Technology\\ Al. Mickiewicza 30, 30-059 Krak\'{o}w, Poland \\
e-mail: kutafina@mat.agh.edu.pl\\[2ex] }
\begin{document}
\maketitle

\begin{abstract}
 We consider the hyperbolic generalization of Burgers equation
  with polynomial source term. The transformation of auto-B\"{a}cklund  type  was found. Application of the results is shown in the examples, where  the pair of two stationary solutions produces   kink and bi-kink solutions.
\end{abstract}
\section{Introduction}
In recent years  efforts of many scientists were  concentrated on
the obtaining exact solutions  for non-integrable PDE's. Special
attention was paid to  solutions describing so-called wave
patterns. A number of interesting results was obtained for
hyperbolic generalization of Burgers equation (GBE):
\begin{equation}\label{gbevol}\tau u_{tt}-\kappa u_{xx}+A\,uu_x+B\,u_t=f(u)=\lambda(u-m_1)(u-m_2)(u-m_3). \end{equation}
 In the papers  \cite{makar1, makar}  equation (\ref{gbevol}) is
derived as a model equation for a generalized Navier-Stokes
system  taking into account the
influence of memory (relaxation) effects.  Thanks to the  constants
$A,\,\,B,\,\,\tau,\,\,\kappa$ GBE can describe a great
amount of special cases e.g. for $A=0$ equation (\ref{gbevol})
coincides with the nonlinear telegraph equation, for
$A=B=0$ -  with  d'Alembert equation, for $\tau=0$ - with Burgers equation etc.. \\

As a nonlinear dissipative equation with outer sources
(\ref{gbevol}) can describe dissipative structures, e.g. soliton-
and kink-like solutions.
 We succeeded in finding out a number of exact and approximated solutions
with the help of combining different known methods like Hirota's
ansatz \cite{hir}, conditional symmetries \cite{WFIT, olver2} and
qualitative analysis \cite{G-H}.  The number of already known
exact solutions to GBE is sufficiently large
\cite{vladku04,  vladku06_sym, vladku06} to pose the problem of
their internal structure and interaction. As it is well known the
superposition rule cannot be applied in nonlinear case. However
for classical Burgers equation
\begin{equation}\label{bg}\kappa u_{xx}-uu_x-u_t=0
\end{equation}
there exists the
  auto-B\"{a}cklund transformation
  \begin{equation}\label{anz_vol}u(x,t)=\frac{M(x,t)Exp(h(x,t))+Q(x,t)}{Exp(h(x,t))+1} \end{equation}
  stating the nonlinear analog to superposition principle. In fact, for
 every pair of solutions $M(x,t)$, $Q(x,t)$ there exists a
 function $h(x,t)$  such that $u(x,t)$ given by (\ref{anz_vol}) is also a
 solution. The transformation (\ref{anz_vol}) is strongly connected with the Cole-Hopf ansatz. V. G. Danilov, V. P. Maslov and
K. A. Volosov \cite{vol, vol06}
 used this ansatz 
  considering  FitzHygh-Nagumo-Semenov equation:
\begin{equation}\label{fns}u_t-\kappa u_{xx}/(2\,a^3)=\kappa (u-u^3), \end{equation}
which is a particular case of (\ref{gbevol}). For this equation
there was shown, that the set of exact solutions of some special
form possesses the structure of a semigroup.\\
 In this paper we make
a step forward by stating the conditions of the existence of auto-B\"{a}cklund
transformation and certain kind of  algebraic structure for some
special sets of solutions to GBE.

 \section{Auto-B\"{a}cklund transformation for GBE}

Let us apply  the ansatz (\ref{anz_vol}) to GBE. Collecting the coefficients at different
powers of $Exp(h(x,t))$ and equating these terms to
zero we obtain four equations: $e_i,\,\,i=0,..,3$:
\begin{equation}\label{ee} e_0: \tau M_{t\,t}(x,t)-\kappa
M_{x\,x}(x,t)+A\,M(x,t)M_x(x,t)+B\,M_t(x,t)-
\end{equation}
$$-\lambda(M(x,t)-m_1)(M(x,t)-m_2)(M(x,t)-m_3)=0$$
$$e_1: 3\lambda\,\,m_1\,m_2\,m_3 + B\,M_{t}(x,t) + 2\,\tau
\,h_{t}(x,t)\,M_{t}(x,t) +
  2\,B\,Q_{t}(x,t) - 2\,\tau \,h_{t}(x,t)\,Q_{t}(x,t)
  +$$
$$  + \tau \,M_{t\,t}(x,t)
  + 2\,\tau \,Q_{t\,t}(x,t) +
  {Q^2(x,t)}\left(\lambda\,( m_1 + m_2 + m_3) - A\,h_{x}(x,t) \right)-$$
 $$  -
  2\,\kappa \,h_{x}(x,t)\,M_{x}(x,t) + 2\,\kappa \,h_{x}(x,t)\,Q_{x}(x,t)-$$
$$ -
  Q(x,t)\,( 2\lambda\,(\,m_1\,m_2 + \,m_1\,m_3 + \,m_2\,m_3) +  B\,h_{t}(x,t) + \tau \,{h_{t}^2(x,t)} + \tau \,h_{t\,t}(x,t) -
   $$
$$ - \kappa \,{h_{x}^2(x,t)} -
     A\,M_{x}(x,t)- A\,Q_{x}(x,t) - \kappa \,h_{x\,x}(x,t))+
  M(x,t)\,( -\lambda\,(m_1\,m_2 + m_1\,m_3+
     m_2\,m_3)- $$
     $$- 3\,\lambda\,Q(x,t)^2 + B\,h_{t}(x,t) + \tau \,{h_{t}(x,t)}^2 +
     \tau \,h_{t\,t}(x,t)  - \kappa \,{h_{x}^2(x,t)}
     +$$
     $$+Q(x,t)\,( 2\lambda\,( m_1 + m_2 + m_3 )  + A\,h_{x}(x,t))  +
     A\,Q_{x}(x,t)-$$
     $$ - \kappa \,h_{x\,x}(x,t))  - \kappa \,M_{x\,x}(x,t) - 2\,\kappa
     \,Q_{x\,x}(x,t)=0$$
$$e_2: 3\lambda\,m_1\,m_2\,m_3+ 2\,B\,M_{t}(x,t) + 2\,\tau \,h_{t}(x,t)\,M_{t}(x,t) +
  B\,Q_{t}(x,t)-$$
  $$ - 2\,\tau \,h_{t}(x,t)\,Q_{t}(x,t) + 2\,\tau \,M_{t\,t}(x,t) + \tau \,Q_{t\,t}(x,t) +
  {M^2(x,t)}\,(\lambda(m_1 + m_2 + m_3 - 3\,Q(x,t))+ $$
  $$ + A\,h_{x}(x,t) ) - 2\,\kappa \,h_{x}(x,t)\,M_{x}(x,t)  +
  2\,\kappa \,h_{x}(x,t)\,Q_{x}(x,t)-$$
  $$ - Q(x,t)\,
  (\lambda( m_1\,m_2+ ( m_1 + m_2 ) \,m_3 )+
     h_{t}(x,t)\,( B - \tau \,h_{t}(x,t) )  + \tau \,h_{t\,t}(x,t)
     + $$
$$ + \kappa \,h^2_{x}(x,t)   -
A\,M_{x}(x,t) - \kappa \,h_{x\,x}(x,t) )  +
  M(x,t)\,( -2\lambda \,( m_1\,m_2 +
        m_1\,m_3 + m_2 \,m_3 )+ $$
        $$ +
     h_{t}(x,t)\,( B - \tau \,h_{t}(x,t) )  + \tau \,h_{t\,t}(x,t) +
     Q(x,t)\,( 2\lambda \,(  m_1 + m_2 + m_3 )-$$
     $$  - A\,h_{x}(x,t) )  +
     h_{x}(x,t)\,(  \kappa \,h_{x}(x,t) )  + A\,( M_{x}(x,t) + Q_{x}(x,t) )  -
     \kappa \,h_{x\,x}(x,t))-$$
     $$- 2\,\kappa \,M_{x\,x}(x,t) - \kappa
     \,Q_{x\,x}(x,t)=0$$
$$e_3: \tau Q_{t\,t}(x,t)-\kappa Q_{x\,x}(x,t)+AQ(x,t)Q_x(x,t)+BQ_t(x,t)-$$
$$-\lambda(Q(x,t)-m_1)(Q(x,t)-m_2)(Q(x,t)-m_3)=0.$$
The key fact for our further considerations is that 
two of them, namely $e_0$ and $e_3$ are  GBE (\ref{gbevol})
written for $M(x,t)$ and $Q(x,t)$. The equations $e_1$ and $e_2$
seem to have quite complicated form. To simplify these equations,
we calculate $M_{x\,x}(x,t)$ and $Q_{x\,x}(x,t)$ from $e_0$ and
$e_3$, put them into $(e_1,\,\,e_2)$ and consider the system
$(e_2-e_1,\,\,e_1)$. Additionally we introduce the new auxiliary
function $P(x,t)=M(x,t)-Q(x,t)$. Finally the system (\ref{ee})
takes the form $(e_0,\,e_3,\,e_4,\,e_5)$, where
\begin{equation}\label{e1-e2} e_4={P^2(x,t)} - 2\,\tau \,{h_{t}^2(x,t)} + A\,P(x,t)\,h_{x}(x,t) + 2\,\kappa \,{h_{x}^2(x,t)}=0,\end{equation}
\begin{equation}\label{e1}e_5={P^3(x,t)}-\lambda\,(m_1+m_2+m_3)\,P^2(x,t) + 3\,{P^2(x,t)}\,Q(x,t) + 2\,\tau \,h_{t}(x,t)\,P_{t}(x,t) -\end{equation}
$$- 2\,\kappa \,h_{x}(x,t)\,P_{x}(x,t) +
 P(x,t)\,( B\,h_{t}(x,t) + \tau \,{h_{t}^2(x,t)} + \tau \,h_{t\,t}(x,t) 
 +$$
 $$+ A\,Q(x,t)\,h_{x}(x,t)
      - \kappa \,{h_{x}^2(x,t)}
     - A\,P_{x}(x,t) - \kappa \,h_{x\,x}(x,t))=0.$$
       Let us note that using the standard scaling transformation $t=e^{\alpha}\,\hat{t},\,\,x=e^\beta\,\hat{x},\,\,u=e^\gamma\,\hat{u}$ one can manipulate some of the parameters. So in further  considerations without loss of the generality we shall assume that $\kappa=1$ while $A=2\sqrt{2}$.  In
this case (\ref{gbevol}) takes form:
 \begin{equation}\label{gbevol_2} \tau\,
u_{tt}+2\sqrt{2}\,uu_x+B\,u_t-
u_{xx}=\lambda (u-m_1)(u-m_2)(u-m_3)
\end{equation}
and $e_4$ becomes linear:
  \begin{equation}\label{e1-e2-2}e_4= \pm\sqrt{ 2\,\tau }\,{h_{t}(x,t)} - P(x,t)- \sqrt{2}\,{h_{x}(x,t)}=0.\end{equation}
  Further on we take into account only "+", since the results for the other possibility are symmetric.
  Solution of (\ref{e1-e2-2}) can be found in
  \cite{evans}:
  \begin{equation}\label{ha}h(x,t)=\frac{1}{\sqrt{2\,\tau}}\int_0^x P(\xi-\frac{t}{\sqrt{\tau}},\, t)d\xi+\Phi(x+\frac{t}{\sqrt{\tau}}),\end{equation} where $\Phi$ is some arbitrary smooth function.
  
 We can formulate the results in the form of following lemma: 
\begin{lemma}\label{vol_1}   If the functions
$M(x,t),\,\,Q(x,t)$ satisfy GBE (\ref{gbevol}), $h(x,t)$ is given by (\ref{ha})  and the
compatibility condition (\ref{e1}) is fulfilled, then
$$u(x,t)=\frac{M(x,t)Exp(h(x,t))+Q(x,t)}{Exp(h(x,t))+1}$$
is the solution to GBE (\ref{gbevol}).
\end{lemma}
The  above lemma shows, that when the equations (\ref{e1},\ref{ha}) are satisfied the pair of solutions to GBE
(\ref{gbevol}) can  produce a new one. Our aim is to define a set of solutions, which pairwise correspond to a certain function $h(x,t)$ and satisfy the compatibility condition (\ref{e1}). Further steps will be as follows:
 first we introduce certain
equivalence relation between the solutions. Then we consider
equivalence class of the constant solution $u(x,t)=m_1$ and inside
this class we build a semi-algebraic structure. The examples of applications are given in the last section.\\

%**************************************************************************
\section{Equivalence relation}
\begin{definition}\label{rel2}Let
$M(x,t)=\frac{M_1(x,t)Exp(h_1(x,t))+Q_1(x,t)}{Exp(h_1(x,t))+1}$,\\
$Q(x,t)=\frac{M_2(x,t)Exp(h_2(x,t))+Q_2(x,t)}{Exp(h_2(x,t))+1}$ be
the pair of solutions to (\ref{gbevol_2})\footnote{Let us notice, that automatically $M_i(x,t)$ and $Q_i(x,t)$, $i=1,2$ must satisfy GBE.}.  $M(x,t)$ and $Q(x,t)$
are said to be in relation
 ( $M(x,t)\sim Q(x,t)$) if and only if  there exists a  function
$h(x,t)$ of the form (\ref{ha}) such that
$M(x,t),\,\,Q(x,t)$ and $h(x,t)$ satisfy  (\ref{e1}).
\end{definition}
\begin{theorem} The relation stated by definition (\ref{rel2}) is an  equivalence relation if $B$ is of the form:
 \begin{equation}\label{b}B=-\frac{  4\,\lambda\, ( {m_1} + {m_2} + {m_3})  \,
    \sqrt{\tau }}{\sqrt{2}}.\end{equation} \end{theorem}
Proof:\\

1. Reflexivity: $M(x,t)\sim M(x,t)$\\
$u(x,t)=\frac{M(x,t)Exp(h(x,t))+M(x,t)}{Exp(h(x,t))+1}=M(x,t)$ is
a GBE solution for  $h(x,t)=0$. \\

2. Symmetry: $M(x,t)\sim Q(x,t)\Leftrightarrow
Q(x,t)\sim M(x,t)$\\

 We assume, that the conditions
(\ref{e1}, \ref{ha}) are satisfied for $P(x,t)=M(x,t)-Q(x,t)$
and $h(x,t)$. Let us find
$\tilde{h}(x,t)$ for $\tilde{P}(x,t)=Q(x,t)-M(x,t)=-P(x,t)$.\\
The condition (\ref{ha}) for ($P(x,t),\,\,h(x,t)$) is
$$\sqrt{ 2\,\tau }\,{h_{t}(x,t)} - P(x,t)-\sqrt{2}\,{h_{x}(x,t)}=0.$$
For ($-P(x,t),\,\,\tilde{h}(x,t)$):
$$\sqrt{ 2\,\tau }\,{\tilde{h}_{t}(x,t)} + P(x,t)-\sqrt{2}\,{\tilde{h}_{x}(x,t)}=0.$$
It is easy to see, that $\tilde{h}(x,t)=-h(x,t)$ is a right choice.
Now we check the second condition  (\ref{e1}) for the pair
($-P(x,t),\,\,-h(x,t)$):
\begin{equation}\label{e1a}-{P(x,t)}^3-\lambda\,(m_1+m_2+m_3)\,P^2(x,t) + 3\,{P(x,t)}^2\,M(x,t) + 2\,\tau
\,h_t(x,t)\,P_t(x,t) \end{equation}
$$- 2\,h_x(x,t)\,P_x(x,t) - P(x,t)\,( -B\,h_t(x,t) + \tau \,{h_t(x,t)}^2 - \tau \,h_{t\,t}(x,t) - $$
    $$-2\sqrt{2}\,M(x,t)\,h_x(x,t)
     - \,{h_x(x,t)}^2 + 2\sqrt{2\,}\,P_x(x,t) +\,h_{x\,x}(x,t)
    )=0.$$
Adding (\ref{e1}) and (\ref{e1a}) results in:

$${P(x,t)}^2 - 2\,\tau \,{h_t(x,t)}^2 + 2\sqrt{2}\,P(x,t)\,h_x(x,t) +
      2\,{h_x(x,t)}^2  =0, $$ which
      is satisfied by assumptions.\\

3. Transitivity: $M_1(x,t)\sim Q_1(x,t),\,\,Q_1(x,t)\sim M_2(x,t)\Rightarrow M_1(x,t)\sim M_2(x,t)$\\
We assume, that ($M_1(x,t),\,\, Q_1(x,t)$) and ($Q_1(x,t),\,\,
M_2(x,t))$
 satisfy (\ref{e1}, \ref{ha}) with respectively
 $h_1(x,t),\,\,h_2(x,t)$. From (\ref{ha}) we can calculate:
 $$P_1(x,t)=M_1(x,t)-Q_1(x,t)= \sqrt{ 2\,\tau }\,{h_{1\,t}(x,t)}-
\sqrt{2}\,{h_{1\,x}(x,t)}$$
$$P_2(x,t)=Q_1(x,t)-M_2(x,t)=\sqrt{ 2\,\tau }\,h_{2\,t}(x,t)-\sqrt{2}\,h_{2\,x}(x,t).  $$
Since we want $M_1(x,t)$ to be in the relation with $M_2(x,t)$
then we need some $h(x,t)$, which satisfies:
$$P(x,t)=M_1(x,t)-M_2(x,t)=\sqrt{2\,\tau}\,h_{t}(x,t)-\sqrt{2}\,h_{x}(x,t).$$
$P(x,t)$ can be written in terms of $h_1(x,t),\,\,h_2(x,t)$:
$$P(x,t)=P_1(x,t)+P_2(x,t)=\sqrt{ 2\,\tau }(h_{1\,t}(x,t)+h_{2\,t}(x,t))-\sqrt{ 2}\,(h_{1\,x}(x,t)+h_{2\,x}(x,t))$$
and we can put
$$h(x,t)=h_1(x,t)+h_2(x,t).$$
In order to check (\ref{e1}) for $h(x,t)$ we use the following
procedure: in the first step we eliminate $h_{1\,x\,x}(x,t)$ and
$h_{2\,x\,x}(x,t)$ from the corresponding versions of (\ref{e1}), next we
use them in (\ref{e1}) for $h(x,t)$, with additional facts that
$M_1(x,t)=P_1(x,t)+Q_1(x,t)$, $M_2(x,t)=P_2(x,t)+Q_1(x,t)$. Simple
but quite cumbersome calculations provide us the additional
condition (\ref{b}).
\nopagebreak[4]
\begin{flushright}\rule{5pt}{5pt}\end{flushright}

\begin{notation}\label{abstr} $\Gamma$ denotes the equivalence class of
the stationary solution $u(x,t)=m_1$ to (\ref{gbevol_2}).
($\Gamma=[m_1]_{\sim}$).\end{notation}
Construction of the set $\Gamma$ implies, that the elements of $\Gamma$ pairwise can produce another solutions. In the next part we introduce some algebraic structure within this set.
This structure is very important for estimation of number of truly new solutions. Let us also note, that $\Gamma$ is not empty since at least it contains the stationary solution $m_1$.

%%%%%%%%%%%%%%%%%%%%%%%%%%%%%%%%%%%%%%%%%%%%%%%%%%%%%%%%%%%%%%%%%%%%%%
\section{Algebraic-like structure}

\begin{definition}\label{op} For a given $h(x,t)$ satisfying (\ref{ha}) we define operation in $\Gamma$:   
$$M(x,t)\circ_h Q(x,t)=\frac{M(x,t)Exp\,(h(x,t))+Q(x,t)}{Exp\,(h(x,t))+1}. $$  \end{definition}
\begin{lemma} \label{close} $\Gamma$ is closed with respect to the operation
"$\circ_h$". In other words $M(x,t),\,Q(x,t)\in\Gamma\Rightarrow
\exists h(x,t):$
 $M(x,t)\circ_h Q(x,t)\in\Gamma.$
\end{lemma}
Proof:\\
Since $M(x,t)\in [m_1]_{\sim}$ and $Q(x,t)\in [m_1]_{\sim}$
the transitivity of relation $\sim$ provides  $M(x,t)\sim Q(x,t)$.
So there exists function $h_1(x,t)$ such that
$u(x,t)=M(x,t)\circ_{h_1} Q(x,t)$ is a  solution of
(\ref{gbevol_2}). We have to check, whether $u(x,t)\in
[m_1]_{\sim}$. In order to do that  we have to find
$h(x,t)$ such that $u(x,t)\circ_{h} m_1$ is also a solution of (\ref{gbevol_2}).\\
Let the pair $(M(x,t),m_1)$ correspond to the function
$h_2(x,t)$, then from the equation (\ref{e1-e2-2}) written for
that  pair we obtain:
$$M(x,t)=\sqrt{2\,\tau}\,h_{2\,t}(x,t)-\sqrt{2}\,h_{2\,x}(x,t)+m_1. $$
The same equation (\ref{e1-e2-2}) written  for the pair
$(M(x,t),Q(x,t))$:
$$M(x,t)-Q(x,t)=\sqrt{2\,\tau}\,h_{1\,t}(x,t)-\sqrt{2}\,h_{1\,x}(x,t) \Rightarrow$$
$$\,\,\,\,\,\,\,\,\,\,\,\,\,\,\Rightarrow
Q(x,t)=\sqrt{2\,\tau}\,(h_{2\,t}(x,t)-h_{1\,t}(x,t))-\sqrt{2}\,(h_{2\,x}(x,t)-h_{1\,x}(x,t))-m_1.$$
Hence the function $h(x,t)$ must satisfy:
$$\sqrt{2\,\tau}\,h_{\,t}(x,t)-\sqrt{2}\,h_{\,x}(x,t)=M(x,t)\circ_{h_1}Q(x,t)-m_1=$$
$$=\frac{M(x,t)Exp(h_1(x,t))+Q(x,t)}{Exp(h_1(x,t))+1}-m_1=\frac{(\sqrt{2\,\tau}\,h_{2\,t}(x,t)-\sqrt{2}\,h_{2\,x}(x,t)+m_1)Exp(h_1(x,t))}{Exp(h_1(x,t))+1}+$$
$$+\frac{\sqrt{2\,\tau}\,(h_{2\,t}(x,t)-h_{1\,t}(x,t))-\sqrt{2}\,(h_{2\,x}(x,t)-h_{1\,x}(x,t))-m_1}{Exp(h_1(x,t))+1}-m_1=$$
$$=\sqrt{2\tau}\left(h_{2\,t}(x,t)-\frac{h_{1\,t}(x,t)}{Exp(h_1(x,t))+1}\right)-\sqrt{2}\left(h_{2\,x}(x,t)-\frac{h_{1\,x}(x,t)}{Exp(h_1(x,t))+1}\right).$$
If we put
\begin{equation}\label{trans}h(x,t)=h_2(x,t)-h_1(x,t)+Ln(1+Exp(h_1(x,t))), \end{equation} then
condition (\ref{e1}) can be easily checked by direct substitution.
\begin{flushright}\rule{5pt}{5pt}\end{flushright}
\begin{theorem}\label{prop}
$\Gamma$ has  the following properties:\\
a) Any element of $\Gamma$ is the unity for itself: $M(x,t)\circ_0 M(x,t)=M(x,t)$\\
b) The operation is commutative: $M(x,t)\circ_h Q(x,t)=Q(x,t)\circ_{-h}M(x,t)$\\
 c) Association property: $M_1(x,t)\circ_{H_1} (Q_1(x,t)\circ_{-h_2} M_2(x,t))=(M_1(x,t)\circ_{h_1} Q_1(x,t))\circ_{H_2}
 M_2(x,t)$, where $H_1(x,t)=-Ln\frac{1+Exp(-h_2(x,t))}{Exp(h_1(x,t))}$ and
$H_2(x,t)=Ln\frac{1+Exp(h_1(x,t))}{Exp(h_2(x,t))}$.

\end{theorem}
Proof:\\
a) The pair $(M(x,t), M(x,t))$ may correspond to the function
$h(x,t)=0$, so:
 $$M(x,t)\circ_0 M(x,t)=\frac{M(x,t)Exp(0)+M(x,t)}{Exp(0)+1}=M(x,t)$$

b) From the theorem \ref{rel2} we know, that if the pair
$(M(x,t),Q(x,t))$ corresponds to $h(x,t)$ then $(Q(x,t),M(x,t))$
is connected with $-h(x,t)$.
$$M(x,t)\circ_h
Q(x,t)=\frac{M(x,t)Exp(h(x,t))+Q(x,t)}{Exp(h(x,t))+1}=$$
$$=\frac{Q(x,t)Exp(-h(x,t))+M(x,t)}{Exp(-h(x,t))+1}=Q(x,t)\circ_{-h}
M(x,t)$$

c) We assume that the pairs $(M_1(x,t),Q_1(x,t))$ and
$(M_2(x,t),Q_1(x,t))$ are connected respectively with $h_1(x,t)$
and $h_2(x,t)$. To find $H_2(x,t)$ we write down (\ref{e1-e2-2})
for the pairs $(M_1(x,t),Q_1(x,t))$, $(M_2(x,t),Q_1(x,t)),$
$(M_1(x,t)\circ (Q_1(x,t)), M_2(x,t))$ and next execute the
calculations similar to  lemma \ref{close}. Finally we obtain:
$$H_2(x,t)=\int\frac{Exp(h_1(x,t))}{Exp(h_1(x,t))+1}h_{1\,x}(x,t)-h_{2\,x}(x,t)dx=Ln\frac{1+Exp(h_1(x,t))}{Exp(h_2(x,t))}.$$
Since the pair $(M_1(x,t)\circ\, Q_1(x,t)), M_2(x,t))$ corresponds
to $H_2(x,t)=Ln\frac{1+Exp(h_1(x,t))}{Exp(h_2(x,t))}$, the
proper function for $(M_1(x,t)\circ\, Q_1(x,t), M_2(x,t))$ would be
$H_1(x,t)=-Ln\frac{1+Exp(-h_2(x,t))}{Exp(h_1(x,t))}$. Direct
substitution finishes the proof.\nopagebreak[4]
\begin{flushright}\rule{5pt}{5pt}\end{flushright}
Let us note, that the  results of present section implies that $(M(x,t)\circ\, Q(x,t))\circ\, Q(x,t)=M(x,t)\circ (Q(x,t)\circ Q(x,t))=M(x,t)\circ Q(x,t)$. It means, that the finite set of the solutions can produce only finite  number of new ones.
%%%%%%%%%%%%%%%%%%%%%f%%%%%%%%%%%%%%5
\section{Examples}
 Let  $Q(x,t)=m_1$, $M(x,t)=m_2$, $\lambda=1$. It is easy to check, that $M(x,t),\,Q(x,t)\in \Gamma$. Then for
$P(x,t)=M(x,t)-Q(x,t)=m_2-m_1$ the equation (\ref{e1-e2-2}) can be
written as follows:
$$ {\sqrt{2\,\tau }}\,h_{\,t}(x,t) - {\sqrt{2}}\,h_{\,x}(x,t)+m_1-m_2=0,$$
so the function $h(x,t)$ can be easily calculated:
$$h(x,t)=\frac{(m_1-m_2)\,x}{{\sqrt{2}}}  + \phi (t + x\,\sqrt{\tau }).$$
The compatibility condition (\ref{e1}) takes form:
\begin{equation}\label{phi}\left( 2\,m_3 + m_2+ m_1 \right)
(m_2-m_1+2\,\sqrt{2\,\tau}\,\phi' (t + x\,\sqrt{\tau}))=0.\end{equation}
  The solutions of the above equation (\ref{phi})
are as follows:
\begin{equation}\label{sol1}\phi (t + x\,\sqrt{\tau})=\frac{(m_1-m_2) \,(t + x\,\sqrt{\tau}) }
  {2 \,\sqrt{2\,\tau } }+c\end{equation} or
\begin{equation}\label{sol2}m_1=-m_2-2\,m_3.\end{equation}
In the first case (\ref{sol1}) function $h(x,t)$ is of the form:
$$h(x,t)=\frac{m_1-m_2}{2\,\sqrt{2\,\tau}}(3\sqrt{\tau}\,x+t)$$ and we obtain a solution of kink type:
\begin{equation}\label{kink1} u(x,t)=\frac{m_2\,Exp(\frac{m_1-m_2}{2\,\sqrt{2\,\tau}}(3\sqrt{\tau}\,x+t))+m_1}{Exp(\frac{m_1-m_2}{2\,\sqrt{2\,\tau}}(3\sqrt{\tau}\,x+t))+1}.\end{equation}
For $m_1=\tau=1$, $m_2=0$ this solution is presented on the figure 1.\\

In the second case (\ref{sol2}) we can  choose smooth function $\phi (t + x\,\sqrt{\tau})$ arbitrary. For example we can take
$$\phi (t + x\,\sqrt{\tau})=Sin(t+ x\,\sqrt{\tau}).$$
Then we obtain a solution of the form: \begin{equation}\label{kink2}u(x,t)=m_2-\frac{2(m_2+m_3)}{1+Exp(-\sqrt{2}\,(m_2+m_3)\,x +Sin(t+ x\,\sqrt{\tau}))}.\end{equation}
Such a solution corresponds to the  periodic kink solution presented on the figure 2 for the parameters $\tau=1,\,\,m_2=2,\,\,m_3=1$.\\  

 For 
$$\phi (t + x\,\sqrt{\tau})=Log \left(\frac{1 + Exp(t + x\,\sqrt{\tau})\,R}{1 + Exp(t + x\,\sqrt{\tau})}\right)$$ we obtain a bi-kink solution: 
\begin{equation}\label{bikink}u(x,t)=\frac{m_2\,Exp(\omega_1)[1+R\,Exp(\omega_2)]}{1+Exp(\omega_1)+Exp(\omega_2)+R\,Exp(\omega_1+\omega_2)},\end{equation}
where
$\omega_1=-\frac{m_2\,x}{\sqrt{2}},\,\,\,\omega_2=t + x\,\sqrt{\tau}.$\\
For $\tau=1,\,\,m_3=-0.1,\,\,R=0.001$ it has the form:
$$u(x,t)=\frac{e^t - 2\,e^x}{e^t + e^{\frac{5\,t}{4} + \frac{x}{2}} +
2\,e^x}$$ and presented
      on the figure 3.
\begin{figure}
\includegraphics[width=70mm, height=50mm]{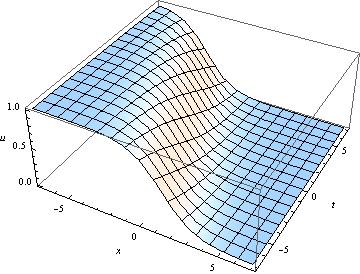}
 \caption{Plot of the solution (\ref{kink1}) for
 $\tau=1,\,\,m_1=1,\,\,m_2=0$}\end{figure}
\begin{figure}\includegraphics[width=70mm, height=50mm]{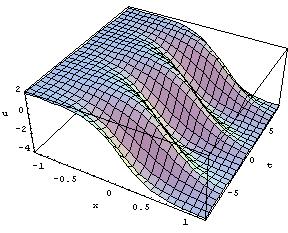}
\caption{Plot of the solution (\ref{kink2}) for
 $\tau=1,\,\,m_2=2,\,\,m_3=1$}\end{figure}
\begin{figure}\includegraphics[width=70mm, height=50mm]{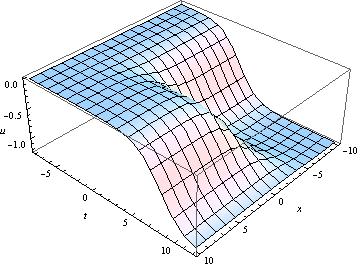}
 \caption{Plot of the solution (\ref{bikink}) for
 $\tau=1,\,\,m_3=-0.1,\,\,R=0.001$ }\end{figure}
\section{Conclusions}
We considered the hyperbolic generalization of Burgers equation (\ref{gbevol}) and
the ansatz (\ref{anz_vol}), which plays role of auto-B\"{a}cklund
transformation for classical Burgers equation. For generalized
equation we  obtained the formula (\ref{ha}) describing
function $h(x,t)$ and the  compatibility condition
(\ref{e1}). The equivalence class $\Gamma$ was constructed in a way allowing to "join" every pair of solutions in a new one. Some algebraic properties of this set was stated. The examples in the last section illustrate the possibility of practical application of these results. Two trivial stationary solutions from the set $\Gamma$ were used
 to obtain a wide class of kink and bi-kink solutions.

\end{document}